# Large Magnon-Induced Anomalous Nernst Conductivity in Single Crystal MnBi

Bin He[1,2,*], Cüneyt Şahin[3,4], Stephen R. Boona[5], Brian C. Sales[6], Yu Pan[1], Claudia Felser[1], Michael E. Flatté [3,4] and Joseph P. Heremans[2,7,8]

1. Max Planck Institute for Chemical Physics of Solids, Dresden, 01187, Germany.
2. Department of Mechanical and Aerospace Engineering, The Ohio State University, Columbus OH, 43210, USA.
3. Pritzker School of Molecular Engineering, University of Chicago, Chicago, Illinois, 60637, USA.
4. Optical Science and Technology Center and Department of Physics and Astronomy, The University of Iowa, Iowa City, Iowa, 52242, USA.
5. Center of Electron Microscopy and Analysis, The Ohio State University, Columbus, OH, 43210, USA.
6. Materials Science and Technology Division, Oak Ridge National Lab, Oak Ridge, TN 37830, USA.
7. Department of Materials Science and Engineering, The Ohio State University, Columbus OH, 43210, USA.
8. Department of Physics, The Ohio State University, Columbus, OH, 43210, USA.
*Correspondence to Bin He (bin.he@cpfs.mpg.de)

## Summary

Thermoelectric modules are a promising approach to energy harvesting and efficient cooling. In addition to the longitudinal Seebeck effect, recently transverse devices utilizing the anomalous Nernst effect (ANE) have attracted interest. For high conversion efficiency, it is required that the material should have a large ANE thermoelectric power and low electrical resistance, the product of which is the ANE conductivity. ANE is usually explained in terms of intrinsic contributions from Berry curvature. Our observations suggest that extrinsic contributions also matter. Studying single-crystal MnBi, we find a very high ANE thermopower (~10 μV/K) under 0.6 T at 80 K, and a transverse thermoelectric conductivity of over 40 A/Km. With insight from theoretical calculations, we attribute this large ANE predominantly to a new advective magnon contribution arising from magnon-electron spin-angular momentum transfer. We propose that introducing large spin-orbit coupling into ferromagnetic materials may enhance the ANE through the extrinsic contribution of magnons.

---

[*] Lead contact: Bin He (bin.he@cpfs.mpg.de)

# Introduction

Thermoelectricity plays a significant role in energy harvesting as well as static cooling applications, with its ability to convert heat directly into electricity and vice versa.[1,2,3,4] Since over 90% of the energy humanity uses today comes from thermal processes, even a very slight improvement in efficiency translates into a large amount of energy saved, for example by recovering the heat wasted in the exhaust of internal combustion engines. To date, most thermoelectric research has focused on the longitudinal Seebeck effect, in which the temperature gradient is parallel to the voltage generated. In this case, the thermopiles have to be connected in series to generate a high voltage. This requires that electrical contacts be made at the hot side of each of the n and p-type elements of each couple, a task that requires the development of a contact technology that minimized thermal diffusion of the contact material into the thermoelectric. The need for individual contacts to each thermocouple elements and the fact that all these contacts are connected in series in the assembled module results in irreversible efficiency loss in contact resistance. Transverse thermoelectric devices, in which the voltage generated is perpendicular to the applied temperature gradient, can avoid these disadvantages. This configuration vastly simplifies the fabrication procedure by making it possible to apply the electrical contact only to a colder side of the thermoelectric material. It also reduces the thermal resistance, where a designated voltage can be generated by simply making the device longer or thicker. The low contact resistance losses of transverse devices compared to Peltier coolers is also an asset on cooling applications. With increasing demand for microdevice cooling, it is essential to achieve large transverse thermoelectric response and reveal the mechanisms to maximize the thermoelectric output power.

Transverse thermoelectrics utilizing the anomalous Nernst effect (ANE) have attracted much attention in recent years.[5,6,7,8,9,10,11] In ANE, the thermoelectric voltage, applied temperature gradient and external field are perpendicular to each other, making it promising for transverse thermoelectric application. ANE can be viewed as the thermal analogue to the anomalous Hall effect (AHE) in magnetic materials. Like the AHE signals, the ANE signal reaches its largest value at the saturation magnetization. ANE has been observed in many magnetic materials, particularly ferromagnets.[5,6,7,8,11] With recent progress in understanding the topology of magnetic materials, it is believed that the net Berry curvature is the origin of intrinsic ANE.[5,6,10,12] A further approach to enhancing the ANE thermopower is to utilize the extrinsic contributions,[13] for instance, magnon contributions. One well-established magnon mediated transport phenomenon is the spin Seebeck effect (SSE),[14,15,16,17,18,19] observed in heterostructure thin films devices. In the SSE, a spin current is thermally excited in a ferromagnetic insulator and injected into a detection layer (Pt) with large spin orbit coupling (SOC), where a transverse voltage is detected via the inverse spin Hall effect (ISHE).[20,21] Inspired by the SSE, we expect a magnonic contribution to the total ANE thermopower when introducing large SOC into magnetic materials, since the SSE and ANE share the identical geometry.

To observe a magnon induced ANE, a large SOC is always required. We decided to focus on magnets with bismuth (Bi), the heaviest stable element with the largest SOC.[22] However, Bi - based ferromagnets have always been rare: Bi has difficulty reacting with 3$d$ metals. To date, MnBi is the only know binary ferromagnetic bismuthide with a high Curie temperature (~630 K),[23] making it potentially a high-temperature permanent magnet. Historically the transport properties of MnBi were long uninvestigated due to the lack of single crystals.[24,25] In 2014, McGuire et al. published a study of single-crystal MnBi that detailed their successful single crystal growth using the flux method.[26] Combining large SOC and ferromagnetism, single crystal MnBi is an ideal system to investigate the magnon induced advective contributions to ANE.

Here, we present our thermomagnetic measurements and the observation of a large ANE signal on two batches of MnBi single crystals, Batch-1 (B1) and Batch-2 (B2). For both batches, we examine two pieces of crystals for both the in-plane (//) and cross-plane (⊥) properties. We find a large ANE that reaches 10 µV/K at 0.6 T and 80 K. Moreover, the anomalous Nernst conductivity reaches over 40 A/Km (higher than any reported value) as a result of low resistivity and large ANE thermopower. By carefully comparing the experimental data with the tight-binding calculations, we confirm that the intrinsic ANE mechanism is not sufficient to explain the observation. We posit that the large ANE may arise from an additional advective transport process induced by magnons: the thermally-driven magnon current may spin-polarize the conduction electrons dynamically, resulting in an additional transverse voltage due to large SOC in MnBi.

**Results**

A scenario of how magnon-electron interactions can contribute to a large ANE is schematically shown in Fig. 1(A). When a temperature gradient is applied to a ferromagnetic material, magnons are excited and can interact with electrons in multiple ways further affecting the thermoelectric transport behavior. In the zero-field case, the magnons can transfer their linear momenta ($\vec{p}$) to the electrons and generate an advective contribution to the Seebeck coefficient, known as magnon-drag thermopower ($S_{MD}$).[27] In the presence of a magnetic field, magnon spin flux can transfer spin-angular momenta ($\vec{S}$) to the itinerant electrons during magnon-electron scattering in the bulk of the FM itself, thereby dynamically spin polarizing the itinerant electrons beyond what is expected from the thermodynamic equilibrium band structure. This corresponds to spin pumping across an interface in the SSE, except that here the spin pumping occurs in the bulk during scattering processes. This electron polarization then can generate a transverse electric field by ISHE: $E_{ISHE} = D_{ISHE}(J_S \times \sigma)$, where $D_{ISHE}$ stands for ISHE efficiency and is determined by the intensity of the SOC. Overall, this additional contribution to the ANE could be labeled a self-SSE term. Since Bi is known to exhibit very large SOC,[22] MnBi should have a large $D_{ISHE}$, giving rise to a potentially large extrinsic contribution to the ANE.

Fig. 1(B) shows the crystal structure and the spin-reorientation (SR) process of MnBi, with Mn as the smaller green sphere and Bi as the larger purple sphere. It crystalizes in hexagonal NiAs structure. The magnetic structure of MnBi is complicated because of its SR process.[26] Below 90 K, the spins are aligned in the *ab*-plane; above it, they start to rotate towards the *c*-axis. By 140 K, SR is completed. Because it can change the normal modes of the magnons, SR has a significant influence on the ANE in MnBi

The resistivities of our four crystals increase with temperature, showing a metallic behavior (Fig. 1(C)). A slight difference in resistivity has been observed from batch to batch, a result of different carrier concentrations induced by the different starting composition for the flux method (See supplemental S1 for details). Such differences have also been observed in the Seebeck coefficients. Fig. 1(D) shows the Seebeck coefficient of the four crystals. The in-plane Seebeck coefficients ($B1_{//}$ and $B2_{//}$) increase as temperature goes up, while the cross-plane thermopowers ($B1_\perp$ and $B2_\perp$) show the opposite trend. The latter behavior is very unusual for a metal, and a tentative explanation is presented along with the Nernst discussion. The measurements on the two pairs of freshly-prepared samples ($B1_{//}$ and $B2_{//}$) and ($B1_\perp$ and $B2_\perp$) indicate good sample-to-sample reproducibility. Meanwhile, thermopowers of B2 crystals are slightly lower than those of the B1, because of slightly higher carrier concentration. Lastly, it should be mentioned that the samples are air-sensitive: the thermoelectric transport properties, for instance, the Seebeck coefficient and ANE thermopower, degrade with repeated thermal cycling as described in the supplementary information (Figs. S3 and S4).

Large ANE signals were observed in both batches of crystals. Figs. 2(A) and (B) show the cross-plane ANE thermopower, $S_{zyx}$, measured on the samples B1⊥ and B2⊥. The largest $S_{zyx}$ is observed on B1⊥ at 80 K, reaching 10 μV/K at 0.6 T. Noting the complex magnetization process at low temperature, the magnetization curve is carefully analyzed and the saturation field is confirmed to be 0.6 T at 80 K by the M-H curve [Figs. S2(a) and (b)]. Such a large ANE thermopower is higher than most reported values,[5,6,7,8,9,11] and is only second to recenly reported compound UCo$_{0.8}$Ru$_{0.2}$Al.[28] The $S_{zyx}$ of B2⊥ reaches ~7 μV/K at 110 K, comparable to that of B1⊥ in the same temperature range. For B1⊥, the $S_{zyx}$ can reach saturation at 1 T from 80 K to 140 K. When the SR process is completed, our crystals do not reach saturation magnetization at 1.4 T, thus no saturation in $S_{zyx}$ is observed on B1⊥ from 180 K to 300 K. At room temperature, our measured ANE thermopower is 2.5 μV/K for B1⊥ and 1.5 μV/K for B2⊥. Figs. 2(C) and (D) show the in-plane ANE thermopower $S_{yxz}$, measured on B1// and B2//. For both samples, a clear saturation field of 0.8 T is observed above 140 K. The largest $S_{yxz}$ values of both crystals are around 2 μV/K at 300 K. A non-linear ANE signal is observed below 140 K on B2//, which possibly originates from a complex magnetic structure induced by the SR process.

Since the origins of intrinsic AHE and ANE are often related to each other, we examined the AHE signals of all four crystals in addition to the ANE. Fig. 3(A) and (B) show the cross-plane AHE resistivities $\rho_{zyx}$ on B1⊥ and B2⊥ at various temperatures. Clear AHE signals have been observed on both samples. The AHE signal of B1⊥ is larger than the signal of B2⊥, which is possibly due to different carrier concentrations and scattering effects. Figs. 3(C) and (D) show the in-plane AHE resistivities $\rho_{yxz}$ on B1// and B2//. Clear AHE signals are observed above 180 K on both samples, with a clear nonlinear $\rho_{yxz}$ on B2// during the SR.

Presently, the net Berry curvature is considered to be the intrinsic mechanism of anomalous transverse transport properties, with skew scattering and side jump as the extrinsic mechanisms of AHE.[29] The net Berry curvature is considered to dominate the good metal region ($10^4$~$10^6$ S/cm), while the extrinsic skew scattering dominates the high conductivity region (>$10^6$ S/cm) and side jump dominates the bad metal region (<$10^4$ S/cm). Since the resistivity of MnBi falls in the good metal region, the Berry curvature should dominate the signals in AHE. Experimentally, we observed a non-linear Hall behavior below 0.3 T in AHE resistivities of the B2 samples [Fig. 3(B) and (D)], which is because of the complex magnetic structure arising from the SR. Additionally, a weak first order phase transition[26] would also affect the magnetic structure. A detailed study of the non-linear Hall effect is beyond the scope of this paper, but can be achieved in the future.

In contrast with the AHE signals, the non-linear ANE signals is only observed on the in-plane measurement [Fig. 2(D)], but not on the cross-plane [Fig. 2(B)]. We believe that the nonlinear signals observed in both in-plane AHE [Fig. 3(D)] and ANE [Fig. 2(D)] share the same origin: the Berry curvature shifting during SR. However, for the cross-plane measurements, the nonlinearity is only found on AHE [Fig. 3(B)] but not ANE [Fig. 2(B)], we therefore speculate that the net Berry curvature is not the dominant mechanism for the ANE. An extrinsic mechanism with a large transverse signal is participating in the ANE. Extrinsic contributions to Nernst effect can arise from interactions between charge carriers and quasi-particles, including magnon-drag,[30] paramagnon-drag,[31,32,33] and phonon-drag.[34] Since magnons have been experimentally detected in MnBi single crystals,[35] we propose they are playing a significant role in thermomagnetic transport.

## Discussion

When the charge carriers are in thermodynamic equilibrium (i.e., in the absence of advective transport processes, such as drag), the ANE is related to the energy derivative of the AHE via the Mott relation in metals and degenerately-doped semiconductors.[36] However, the Mott relation breaks down in the presence of drag contributions because it assumes an electron energy distribution that follows equilibrium Fermi-Dirac statistics.[36,37] Under drag conditions, the electron population, taken in isolation, is not at equilibrium. Since the Mott relation permits the calculation of an intrinsic contribution to the ANE that has the same origin as that of the AHE, we first derive the tensor elements of the experimental thermoelectric conductivity tensor $\overleftrightarrow{\alpha}$ at various temperatures. This tensor $\overleftrightarrow{\alpha}$ is related to the thermopower tensor $\overleftrightarrow{S}$ and the conductivity tensor $\overleftrightarrow{\sigma}$, and in particular, the anomalous Nernst conductivity $\alpha_{xy} = S_{xx}\sigma_{yx} + S_{xy}\sigma_{xx}$.

Fig. 4(A) shows the temperature dependent ANE thermopowers from 80 K to 300 K. The in-plane ANE thermopower ($S_{yxz}$) of both batches increased with temperature, whereas the cross-plane ANE thermopower ($S_{zyx}$) decreased monotonically with temperature, from a maximum of 10 μV/K at 80 K to about 2 μV/K at room temperature. Fig. 4(B) shows the anomalous Hall conductivities (AHCs) on B1$_\perp$ and B1$_{//}$. (Note that we are particularly interested in B1 because of the larger ANE signals.) The cross-plane AHC is approximately 800 S/cm below 140 K, while above 140 K it decreases with temperature because of non-saturation behavior. The in-plane AHC increases with temperature, reaching ~ 200 S/cm at 300 K. The AHCs in both directions fall in the intrinsic region of the AHE, which is consistent with the analysis on the longitudinal electrical conductivity.

With the acquired AHCs, we calculated the in-plane and cross-plane thermoelectric linear response tensor elements ($\alpha_{//}$ and $\alpha_\perp$), and compared them with tight-binding results for B1 in Fig. 4(C). Experimentally, $\alpha_\perp$ decreased with temperature, with the largest value of ~44 A/Km at 80 K. Such a large transverse thermoelectric conductivity is an order of magnitude higher than known ferromagnets and antiferromagnets.[5,6,7,8] In theory, both the intrinsic $\alpha_{//}$ and $\alpha_\perp$ should increase with temperature (shown in the embedded figure), and have the absolute values between 0 and 1 A/Km, much smaller than the experimental results. In addition, we examined the $\alpha_\perp$ with the Fermi energy shifting from -5eV to 5eV, however, the absolute value of $\alpha_\perp$ never exceeds 1 A/Km (Fig. S5). Thus, the intrinsic contribution does not by itself explain the behavior of the $\alpha_\perp$ sufficiently; magnon mediated transport can cause a much larger transverse thermoelectric response. In a magnon mediated transport process, large ANE thermopower and longitudinal conductivity are able to coexist because magnons are Bosons, and they are not subjected to the Fermi Dirac distribution. The large ANE is related to the magnon population and magnon electron interaction, not to the electronic band structure. With a large ANE thermopower and a low resistivity, we achieve a giant $\alpha_{ANE}$. The observed temperature dependence gives further evidence for this interpretation. At low temperature, the local spins are aligned in the *ab*-plane, and $\Delta T$ is applied along the *c*-axis. The temperature gradient can excite the magnons and give rise to the self-SSE after the domains are aligned by an external field. In this case, self-SSE can be the origin of the giant Nernst thermopower in B1$_\perp$. Between 90 K and 140 K, the local spins start to rotate from *ab*-plane to *c*-axis, and the magnon dispersion relation also begin to change. The spin orientation reduces the total number of magnons excited by the temperature gradient along the *c*-axis direction, so the self-SSE signal should decrease with temperature. We show the temperature normalized ANE thermopower in Fig. S8, which is supportive to the proposal of low temperature magnon-drag effect.

This advective picture is also consistent with the fact that the longitudinal cross-plane thermopower decreases with increasing temperature (Fig. 1(C)), assuming a large magnon-drag component. According to a previous study,[27] in a simple model the magnon drag thermopower and magnon drag ANE are related to each other by the formula $S_{ANE} = C\mu_0 H \frac{S_{md}}{\rho}$, where *C* is a material parameter depending on the effective mass and scattering mechanism, $\mu_0 H$ is the applied field, $S_{md}$ is the magnon drag thermopower and $\rho$ is the resistivity. Experimentally, both longitudinal and transverse thermopowers decrease with temperature, which agrees qualitatively with the prediction of the formula above. Such a decreasing trend is indicative of the magnon drag contribution to both longitudinal and transverse thermoelectric response. Lacking the detailed band parameters and scattering parameters, we are unable at this time to further quantify the value of C.

Finally, in Fig. 4(D) we compare our experimental ANE thermopower $S_{ANE}$ and transverse thermoelectric conductivity $\alpha_{ANE}$ with other magnetic materials with large ANE responses.[5,6,7,8,9,28] MnBi's $S_{ANE}$ is about 25% higher than that of $Fe_3Ga$ and $Co_2MnGa$, for both of which the intrinsic contributions are recognized as the origin of the large ANE signal. More importantly, the anomalous Nernst conductivity of MnBi is outstanding among all magnetic materials, reaching over 40 A/Km. Even compared to $UCo_{0.8}Ru_{0.2}Al$ with a larger ANE thermopower, our experimental $\alpha_{ANE}$ is still three times higher than that of $UCo_{0.8}Ru_{0.2}Al$. With the extrinsic magnon-drag contribution, MnBi can have the large ANE thermopower while maintaining a low longitudinal resistivity. Such a large $\alpha_{ANE}$ suggests that this magnon drag induced spin angular momentum transfer procedure is a highly effective approach to generating a large transverse thermoelectric response, which can be a new strategy for high-performance themoelectric applications.

In summary, MnBi, with a large spin orbit coupling as well as strong ferromagnetism, is an ideal candidate for studying magnon mediated transport phenomena, and therefore achieving large transverse thermoelectric response. A large ANE signal of 10 µV/K and a record anomalous Nernst conductivity of over 40 A/Km were observed in single crystal MnBi at 80 K. This giant transverse thermoelectric response arises from an extrinsic magnon-electron interaction process. The magnon-electron spin angular momentum transfer process significantly enhances the ANE signal, equivalent to a self-spin Seebeck effect. Our study provides a new fundamental understanding of ANE, which can be quite large in ferromagnets with strong spin orbit interaction. Such utilization of the magnon-electron interactions provides routes for the enhancement of ANE, and will doubtless find applications, as well as be generalized to other topics in the thermoelectrics field.

## Experimental Procedure

### Resource Availability

**Lead Contact**

Further information and requests for resources and materials should be directed to and will be fulfilled by the lead contact, Bin He (bin.he@cpfs.mpg.de).

**Materials Availability**

This study did not generate new unique materials.

**Data and Code Availability**

All data from this study are available from the lead contact upon reasonable request.

Single crystals of MnBi were grown at Oak Ridge National Lab (ORNL) using the method detailed in McGuire et al;[26] they are called Batch-1 (B1) in this paper. The crystals grown at Max Plank Institute for Chemical Physics of Solids (CPfS) were grown in the same method with a slightly higher starting Mn molar fraction (9%); they are called Batch-2 (B2). Optical image, Laue diffraction result and other details is reported in supplementary information part 1, Fig. S1 and Table S1. MnBi single crystals grow as hexagonal cylinders. We measured the transport properties in a modified Janis liquid nitrogen flow cryostat, as well as a Quantum Design Physical Properties Measurement System with a breakout box. In both measurement systems, the set-ups were identical as described previously.[38] We report our result on B1$_{//}$ and B2$_{//}$ for the in-plane transport properties and B1$_\perp$ and B2$_\perp$ for the cross-plane properties. The temperature dependent magnetization is measured in a Quantum Design Magnetic Properties Measurement System 3 from 80 K to 300 K, up to 5 T.

We use the following notation for the transverse transport properties, which are denoted by indices $xyz$. Here, $x$ is the direction of the applied thermodynamic flux (charge or heat flux), $y$ is the direction of the measured voltage, and $z$ is the field direction. Thus, $S_{yxz}$ is the in-plane Nernst thermopower with the magnetic field along the $z$ [0001] axis and measured on crystal B1$_{//}$ / B2$_{//}$, while $S_{zyx}$ is the cross-plane Nernst thermopower with the field parallel to the $a$-axis [$2\bar{1}\bar{1}0$] in the hexagonal lattice, measured on crystals B1$_\perp$/B2$_\perp$. The sign convention in this experiment is opposite to the Gerlach sign conventions.

We constructed a tight-binding Hamiltonian using the parameters derived from the Density

Function Theory.[39] MnBi has a point group of $D_{6h}$ and a space group of $P6_3$, with a hexagonal crystal structure and 4 atoms per unit cell. The lattice constants were taken as $a=b=4.285$ Å and $c=6.113$ Å. The tight-binding Hamiltonian consisted of $p$-orbitals of Bi and $d$-orbitals of Mn. Magnetism was incorporated into the Hamiltonian through the Stoner formalism with parameters 4.5 and 0.2 eV for $d$- and $p$-orbitals, respectively. We also have added the spin-orbit Hamiltonian with spin-orbit couplings of 0.048 and 1.4 eV for Mn and Bi, respectively. The Berry curvature $\Omega_{ij}$, which is an intrinsic property of the electronic band structure, was computed from this 16-band tight-binding Hamiltonian by:

$$\Omega_{ij}(n\mathbf{k}) = Im \sum_{n \neq n'} \frac{<u_{n\mathbf{k}}\left|\frac{\partial H}{\partial k_i}\right|u_{n'\mathbf{k}}><u_{n'\mathbf{k}}\left|\frac{\partial H}{\partial k_i}\right|u_{n\mathbf{k}}>}{(\varepsilon_n - \varepsilon_{n'})^2}$$

For zero magnetization (or no external magnetic field) d-orbitals are located densely around 0 eV energy. As magnetization is increased with applied field, bands that predominantly consist of d-orbitals move away from the center to higher and lower energies. The energy-resolved AHC is calculated by integrating the Berry curvatures

$$\sigma_{ij} = -\frac{e^2}{\hbar}\int d\mathbf{k} \sum_n \Omega_{ij}(n\mathbf{k}) f_{n\mathbf{k}}$$

where $\Omega_{ij}$ is the Berry curvature, and the summation is performed over all occupied bands in the first Brillouin zone. The intrinsic contribution to the ANC is related to the sum of the Berry curvatures and can be calculated by the well-established relation:[40]

$$\alpha_{ij} = -\frac{1}{e}\int d\varepsilon \frac{\partial f(\varepsilon)}{\partial \mu}\sigma_{ij}(\varepsilon)\frac{\varepsilon - \mu}{T}$$

where $e$ is the electric charge, $f$ is the Fermi-Dirac distribution, $T$ is the temperature, $\sigma$ denotes the energy resolved intrinsic AHC, and $\mu$ is the chemical potential.

## Acknowledgements

B.H. and J.P.H. acknowledge support from the the Center for Emerging Materials, an NSF MRSEC grant (DMR-2011876). B.H., Y.P. and C.F. acknowledge support from ERC TOPMAT, the European Union (grant No. 742068) and European Union's Horizon 2020 research and innovation program (grant No. 766566). C.Ş. and M.E.F. acknowledge support from the Center for Emergent Materials, an NSF MRSEC under Award No. DMR-1420451. B.C.S. acknowledge support from the U.S. Department of Energy, Office of Science, Basic Energy Sciences, Materials Sciences and Engineering Division.

## Author Contributions

Conceptualization, B.H., S.R.B., and J.P.H.; Experiments, B.H., S.R.B., B.C.S., and Y.P; Calculations, C.S. and M.E.F.; Analysis and Discussion, B.H., C.S., S.R.B., Y.P., C.F., M.E.F., and J.P.H.; Writing and Revision, B.H., C.S., S.R.B., B.C.S., Y.P., C.F., M.E.F., and J.P.H.

## Declaration of Interests

The authors declare no competing interests.

## References

1. Goldsmid, H. J. (2010). *Introduction to Thermoelectricity* (Springer).
2. Bell, L. (2008). Cooling, Heating, Generating Power, and Recovering Waste Heat with Thermoelectric Systems. *Science* **321**, 1457-1461.
3. Snyder, G. J., and Toberer, E.S. (2008). Complex thermoelectric materials. *Nat. Mater*. 7, 105–114.
4. Heremans, J. P., Dresselhaus, M., Bell, L. E., and Morelli, D.T. (2013). When thermoelectrics reached the nanoscale. *Nat. Nanotech*. 8, 471–473.
5. Sakai, A., Minami, S., Koretsune, T., Chen, T., Higo, T., Wang, Y., Nomoto, T., Hirayama, M., Miwa, S., Nishio-Hamane, D., et al. (2020). Iron-based binary ferromagnets for transverse thermoelectric conversion. *Nature* **581**, 53–57.
6. Sakai, A., Mizuta, Y. P., Nugroho, A. A., Sihombing, R., Koretsune, T., Suzuki, M-T., Takemore, N., Ishii, I., Nishio-Hamane, D., Arita, R., et al. (2018). Giant anomalous Nernst effect and quantum critical scaling in a ferromagnetic semimetal. *Nat. Phys*. **14**, 1119-1124.
7. Guin, S., Vir, P., Zhang, Y., Kumar, N., Watzman, S. J., Fu, C., Liu, E., Manna, K., Schnelle, W., Gooth, J., et al. (2019). Zero-field Nernst effect in a ferromagnetic Kagome-lattice Weyl-semimetal $Co_3Sn_2S_2$. *Adv. Mater*. **31**, 186022.
8. Guin, S. N., Manna, K., Noky, J., Watzman, S. J., Fu, C., Kumar, N., Schnelle, W., Shekhar, C., Sun, Y., Gooth, J., and Felser, C. (2019). Anomalous Nernst effect beyond the magnetization scaling relation in the ferromagnetic Heusler compound $Co_2MnGa$. *NPG Asia Mater*. **11**, 16.
9. Ikhlas, M., Tomita, T., Koretsune, T., Suzuki, M.-T., Nishio-Hamane, D., Arita, R., Otani, Y., and Nakasuji, S. (2017). Large anomalous Nernst effect at room temperature in a chiral antiferromagnet. *Nat. Phys*. **13**, 1085-1090.
10. Liang, T. Lin, J., Gibson, Q., Gao, T., Hirschberger, M., Liu, M., Cava, R. J., and Ong, N. P. (2017). Anomalous Nernst Effect in the Dirac Semimetal $Cd_3As_2$. *Phys. Rev. Lett*. **118**, 136601.

11. Ramos, R., Aguirre, M. H., Anadón, A., Blasco, J., Lucas, I., Uchida, K., Algarabel, P. A., Morellón, L., Saitoh, E., and Ibarra, M. R. (2014). Anomalous Nernst effect of Fe3O4 single crystal. *Phys. Rev. B* **90**, 054422

12. Sykora, S., Caglieris, F., Wuttke, C., Büchner, B., and Hess, C. (2018). Giant anomalous Nernst effect in Weyl semimetals TaP and TaAs. *Phys. Rev. B* **98**, 201107(R).

13. Papaj, M., and Fu, L. (2021). Enhanced anomalous Nernst effect in disordered Dirac and Weyl materials. *Phys. Rev. B* **103**, 075424.

14. Uchida, K., Takahashi, S., Harii, K., Ieda, J., Koshibae, W., Ando, K., Maekawa, S., and Saitoh, E. (2008). Observation of the spin Seebeck effect. *Nature* **455**, 778–781.

15. Uchida, K., Adachi, H., An, T., Ota, T., Toda, M., Hillebrands, B., Maekawa, S., and Saitoh, E. (2010). Spin Seebeck insulator. *Nat. Mater.* **9**, 894–897.

16. Jaworski, C. M. et al. (2012). Giant spin Seebeck effect in a non-magnetic material. *Nature* **487** 210-213.

17. Uchida, K. et al. (2011). Long-range spin Seebeck effect and acoustic spin pumping. *Nat. Mater.* **10**, 737-741.

18. Jaworski, C. M. et al. (2011). Spin-Seebeck effect: A phonon driven spin distribution. *Phys. Rev. Lett.* **106**, 186601.

19. Jaworski, C. M., Myers, R. C., Johnston-Halperin, E., and Heremans, J. P. (2010). Observation of the spin-Seebeck effect in a ferromagnetic semiconductor. *Nat. Mater.* **9**, 989-902.

20. Bauer, G. E. W., Saitoh, E., and van Wees, B. J. (2012). Spin caloritronics. *Nat. Mater*. 11, 391–399.

21. [Boona], S. R., Myers, R. C., and Heremans, J. P. (2014). Spin caloritronics. *Energy Environ. Sci*. 7, 885–910.

22. Heremans, J. P., Cava, R. J., and Samarth, N. (2017). Tetradymites as thermoelectrics and topological insulators. *Nat. Rev. Mater.* **2**, 17049.

23. Adams, E., Hubbard, W. M., and Syeles, A. M. (1952). A new permanent magnet from powdered manganese bismuthide. *J. Appl. Phys.* **23**, 1207.

24. Poudyal, N., and Liu, J. P. (2013). Advances in nanostructured permanent magnets research. *J. Phys. D: Appl. Phys.* **46**, 043001.

25. Williams, H. J., Sherwood, R. C., and Boothby, O. L. (1957). Magnetostriction and magnetic anisotropy of MnBi. *J. Appl. Phys.* **28**, 445.

26. McGuire, M. A., Cao, H., Chakoumakos, B. C, and Sales, B. C. (2014). Symmetry-lowering lattice distortion at the spin reorientation in MnBi single crystals. *Phys. Rev. B* **90**, 174425.

27. Watzman, S. J., Duine, R. A., Tserkovnyak, Y., Boona, S. R., Jin, H., Prakash, A., Zheng, Y., and Heremans, J. P. (2016). Magnon-drag thermopower and Nernst coefficient in Fe, Co, and Ni. *Phys. Rev. B* **94**, 14.

28. Asaba, T., Ivanov, V., Thomas, S. M., Savrasov, S. Y., Thompson, J. D., Bauer, E. D., and Ronning, F. (2021). Colossal anomalous Nernst effect in a correlated noncentrosymmetric kagome ferromagnet. *Sci. Adv.* **7**, 13 eabf1467.

29. Nagaosa, N., Sinova, H., Onoda, S, MacDonald, A. H., and Ong, N. P. (2010). Anomalous Hall effect. *Rev. Mod. Phys.* **82**, 1539-1592.

30. Blatt, F. J., Flood, D. J., Rowe, V., Schroeder, P.A., and Cox, J. E. (1967). Magnon-drag thermopower in iron. *Phys. Rev. Lett.* **18** 395.

31. Zheng, Y., Lu, T., Polash, M.H., Rasoulianboroujeni, M., Liu, N., Manley, M. E., Deng, Y., Sun, P. J., Chen. X. L., Hermann, R. P., et al. (2019). Paramagnon-drag in high thermoelectric figure of merit Li-doped MnTe. *Sci. Adv.* **5** eaat9461. DOI: 10.1126/sciadv.aat9461

32. Ahmed, F., Tsuji, N., and Mori, T. (2017). Thermoelectric properties of $CuGa_{1-x}Mn_xTe_2$: power factor enhancement by incorporation of magnetic ions. *J. Mater. Chem. A* **5**, 7545-7554.

33. Vaney, J. B., Aminorroaya Yamini, S., Takaki, H., Kobayashi, K., Kobayashi, N., and Mori, T. (2019). Magnetism-mediated thermoelectric performance of the Cr-doped bismuth telluride tetradymite. *Mat. Today Phys.* **9,** 100090.

34. Ziman, J. M. (1960). *Electrons and Phonons*. Clarendon Press, Oxford.

35. Williams, T. J., Taylor, A. E., Christianson, A. D., Hahn, S. E., Fishman, R. S., McGuire, M. A., Sales, B. C., and Lumsden, M. D. (2016). Extended magnetic exchange interactions in the high-temperature ferromagnet MnBi. *Appl. Phys. Lett.* **108**, 192403.

36. Cutler, M., and Mott, N. F. (1969). Observation of Anderson localization in an electron gas. *Phys. Rev.* **181,** 1336.

37. Heremans, J. P. (2020). Thermal spin transport and spin in thermoelectrics. arXiv:2001.06366, *La Rivista del Nuovo Cimento* (accepted, 2020).

38. Heremans, J. P., Thrush, C. M., and Morelli, D. T. (2004). Thermopower enhancement in lead telluride nanostructures. *Phys. Rev. B* **70**, 115334.

39. Shanavas, K. V., Parker, D., and Singh, D. J. (2014). Theoretical study on the role of dynamics on the unusual magnetic properties in MnBi. *Sci. Rep.* **4**, 7222.

40. Xiao, D., Yao, Y., Fang, Z., and Niu, Q. (2006). Berry-phase effect in anomalous thermoelectric transport. *Phys. Rev. Lett.* **97**, 026603.

## Main Figure Titles and Legends

**Figure 1: Overview of the experiment and longitudinal transport properties.** (A) Schematic drawing of the magnon induced advective transport. When a temperature gradient is applied to a ferromagnet, it is transferred from the phonon system to the magnon system. The magnons can then carry linear momentum ($\vec{p}$) and spin angular momentum ($\vec{S}$). Transfer of ($\vec{p}$) generates the magnon-drag thermopower $S_{MD}$. In a material with large SOC, transfer of $\vec{S}$ creates an out-of-equilibrium additional spin polarization of the conduction electrons that can generate a transverse thermoelectric voltage via the inverse spin Hall effect; this contributes to the ANE. (B) Crystal structure of MnBi, with the preferential spin orientation labelled in red arrow. From 90 K to 140 K, the spins reorient from *ab*-plane to *c*-axis. (C) Resistivity of four crystals $B1_{//}$ – $B2_{\perp}$ measured from 80 K to 300 K, with B1 and B2 stand for Batch-1 and Batch-2 crystals. The sample-to-sample variation in resistivity is possibly from different carrier concentration. (D) Seebeck coefficient of four samples $B1_{//}$ – $B2_{\perp}$. In-plane thermopowers have a positive temperature dependence while the cross-plane thermopowers show a negative temperature dependence, which we speculate is related to the $S_{MD}$. Again, due to different carrier density, B1 samples have higher thermopower than B2 samples.

**Figure 2: Field-dependent Anomalous Nernst thermopower of MnBi along different directions.** Cross-plane ANE $S_{zyx}$ on samples (A) $B1_{\perp}$ and (B) $B2_{\perp}$, with applied field parallel to *a*-axis, and in-plane ANE $S_{xyz}$ on samples (C) $B1_{//}$ and (D) $B2_{//}$, with applied field parallel to *c*-axis. Measurements were taken down to the 90 K for $B2_{//}$, below which the magnet anisotropy teared apart the sample. On sample $B2_{\perp}$, due to the small size of the crystal (~ 1.2 mm), we are able to create measurable temperature gradient at 110 K. For $S_{xyz}$, the average temperature rise is ~ 7 K with the sink temperature is at 110 K. At 100 K, the average temperature rise is over 10 K when a stable measurable temperature gradient is created. For this reason, the thermoelectric transport measurements are terminated at 110 K for B2$_\perp$. For the longer B1$_\perp$, we are able to measure down to 80 K without overheating the sample. The complete field dependent Nernst thermopower is shown in Fig. S6, from which a clear sign change of the ordinary Nernst signal is observed.

**Figure 3: Anomalous Hall effect in MnBi.** Cross-plane anomalous Hall resistivities of two cross-plane samples measured on (A) B1$_\perp$ and (B) B2$_\perp$, with the field applied along *a*-axis, and in-plane AHE resistivities measured on samples (C) B1$_{//}$ and (D) B2$_{//}$, with applied field parallel to *c*-axis. Nonlinear signals are detected on the field sweep measurement of B2 sample, possibly because of complex magnetic structure induced by spin reorientation and a weak first order phase transition. We show the complete Hall curves in the supplementary information (Fig. S7), from which the ordinary Hall signal is resolved.

**Figure 4: Giant transverse thermoelectric conductivity in MnBi single crystal.** (A) Temperature dependence of anomalous Nernst thermopower. For in-plane direction, data are taken at 0.8 T. For cross-plane direction, data are taken at saturation field below 140 K, and 1 T above 140 K. (B) Anomalous Hall conductivity calculated with measured longitudinal and transverse resistivities. Points are calculated $\sigma_{AHE}$; curves are guide to eye. (C) Theoretical and experimental transverse thermoelectric linear response tensor element $\alpha_{yz}$ and $\alpha_{xy}$, comparing with tight-binding calculated by the tight-binding Hamiltonian. (D) Comparison of the experimental $\alpha_{yz}$ and $S_{yz}$ with other magnetic materials. The giant $\alpha_{yz}$ is a result of low resistivity and large anomalous Nernst thermopower. The anomalous Nernst conductivity of MnBi is outstanding compared to other magnetic materials because of the magnon mediated transport signature. The dashed lines are guide to eye.